\documentstyle[preprint,eqsecnum,aps]{revtex}
\begin{document}

\draft
\title {\Large\bf  PARTICLE PRODUCTION AND CLASSICAL CONDENSATES 
IN DE SITTER SPACE}

\author{Milan Miji\'c\footnote{e-mail address: milan@moumee.calstatela.edu}}
\address{Department of Physics and Astronomy, California State University, 
Los Angeles, California 90032}
\address{ and Institute for Physics, P.O. Box 522, 11001, 
Belgrade, Yugoslavia}

\maketitle

\begin{abstract}
The cosmological particle production in a $k=0$ expanding de Sitter universe 
with a Hubble parameter $H_0$ is considered for various values of mass
or conformal coupling of a free, scalar field. One finds that, for a 
minimally coupled field with mass
$0 \leq m^2 < 9 H_0^2/4$ (except for $m^2 = 2H_0^2$), the one-mode occupation
number grows to unity soon after the physical wavelength of the mode
becomes larger than the Hubble radius, and afterwards diverges as
$n(t) \sim O(1) (\lambda_{phys}(t)/H_0^{-1})^{2\nu}$, where 
$\nu \equiv [9/4 - m^2/H_0^2]^{1/2}$.
However, for a field with $m^2 > 9H_0^2/4$, the occupation number of a mode
outside the Hubble radius is rapidly oscillating and bounded and does 
not exceed unity. These results, readily generalized for cases of a 
nonminimal coupling, provide a clear argument that the long-wavelength 
vacuum fluctuations of low-mass fields in an inflationary 
universe do show classical behavior, while those of heavy fields do not.
The interaction or self-interaction does not appear necessary
for the emergence of classical features, which are entirely due to the
rapid expansion of the de Sitter background and the
upside-down nature of quantum oscillators for modes outside the Hubble
radius. 
\end{abstract}
\pacs{98.80.Cq, 04.62.+v}
\narrowtext

\section{Introduction}

Our current understanding of the formation of structures in the Universe
rests on the remarkable evolution of vacuum fluctuations during the inflationary
phase. While stretched by the rapid expansion from microscopic
to astronomical scales, these fluctuations supposedly
decohere, and start to behave as (the source of) classical density 
perturbations. 

To date, several physical pictures and formalisms have
been put forward as a means to understand this \lq\lq quantum
to classical transition.\rq\rq $~$ Most authors approached the problem
within the general framework of decoherence of a subsystem coupled
to the environment. Two early applications of this idea to density 
perturbations in inflationary cosmology are
those of Sakagami \cite {Sakagami}, and Brandenberger et al. \cite {BLM}.
Unruh and Zurek \cite{UnrZur} examined the classical transition in the case of 
a scalar field coupled to a thermal bath. The general criteria for decoherence
and the study of cosmological perturbations in the squeezed state formalism 
were given by Albrecht et al. \cite{Albrecht}.
Hu and Calzetta \cite {HuCal} introduced the technique of a 
closed time path for evaluating the decoherence functional, and studied in 
depth the mutual influence of particle production, dissipation, 
and decoherence. The same method was applied to the problem of density 
perturbations by  Hu et al. \cite {HuPZ}. Other related works are those of 
Calzetta and Mazzitelli \cite {CalMaz}, who considered decoherence due to 
particle production, Calzetta and Hu \cite {CalHu} who studied the decoherence 
of a mean field by its quantum fluctuations, and Calzetta and Gonorazky \cite
{CalGon}, who investigated fluctuations in a model with non-linear 
coupling. A phenomenological model for the decoherence of density
perturbations was constructed by Kubotani et al. \cite{Morikawa}.

Another category of decoherence studies in inflationary cosmology starts with 
the pioneering work of Guth and Pi \cite {GuthPi}, who showed that at 
late times the wave functions for long-wavelength modes behave as classical 
Gaussian probability distributions. Around the same time Lyth \cite{Lyth} 
pointed out that the wave packet for massless modes in de Sitter space does
not spread, and the quantum uncertainty becomes negligible. More recently,
Polarski and Starobinsky \cite {PolStar} and Lesgurges et al. \cite {LPS}
examined particle production in de Sitter space and found a strong
squeezing of long wavelength modes which makes their behavior essentially
classical. In all of these cases one looks at the free fields, and the 
classical evolution emerges due to a particular property of the de Sitter 
background. Polarski and Starobinsky \cite {PolStar} call this
\lq\lq decoherence without decoherence,\rq\rq $~$ to stress that classical
features may emerge even in the absence of an interaction which is deemed essential in usual decoherence studies. 

A similar but complementary line of reasoning emerges within the stochastic 
approach to inflation \cite {StarMeudon} which dramatically exhibits the
influence of quantum fluctuations on semiclassical evolution. Starobinsky 
\cite{Star82} already pointed out the negligible canonical commutator of long 
wavelength modes. Nambu \cite {Nambu} constructed a statistical operator 
for the coarse-grained field, and argued that its evolution is effectively
classical. In a different picture, Ref. \cite{sdcgqf} demonstrated the 
random walk of a highly peaked wave function for a coarse-grained field, in
accord with the Langevin equation. Finally, Matacz \cite {Matacz} constructed
a new formulation of stochastic inflation and investigated the resulting
semiclassical evolution.

All these different physical pictures provide much welcomed complementary 
insights into the problem, and often have vastly different powers in explaining 
slightly modified or more general situations. Some differences in the results
and interpretations should not be overemphasized at this time. We are still in 
the phase of collecting new evidence through studies of different models, and 
testing of various decoherence criteria and their comparison.

In this spirit, I will aprroach here
the decoherence of vacuum fluctuations in De Sitter space using
the most traditional technique of quantum field theory in curved spacetime:
particle production in an expanding universe \cite {partprod}. One elementarty
criteria for the semiclassical behavior is to look at the magnitude of the
particle occupation number. When this number is large, the canonical coordinate and momentum operators approximately commute, and the system behaves
semiclassicaly. Our goal here is to investigate the decoherence of vacuum 
fluctuations in De Sitter space according to this criteria. Our basic model 
consists of fluctuations of a massive scalar field. We will establish how 
decoherence depends on the mass or conformal coupling of the scalar field.  

Calculations of particle production usually deal with comparing the particle
number at asymptotically early and late times, or with respect to vacuum
states defined in two different frames. In the de Sitter case, the calculation
of the latter kind has been done by Lapedes \cite {Lapedes} who in this way
rederived the existence of a thermal bath seen by an observer in the static 
frame \cite {GibHaw}. In contrast, a different quantity is calculated
here: the time dependent, one-mode occupation number in a spatially
flat, expanding frame. This will allows us to follow the eventual buildup
of (quasi)particle occupation number in the course of the cosmological expansion.

The method to do so is not new, but apparently it has not been 
applied before to the De Sitter case. The probable reason is that since
the pioneering works in Ref. \cite {fluctuations} most efforts 
concerning fluctuations in inflationary cosmology have 
been to compute the spectrum and the evolution of density perturbations in 
various models, and to confront these results with the observational
constraints on the microwave background anisotropy and large scale structure. 
The classical nature of fluctuations is usually assumed, or argued for
in passing, by invoking Gibbons-Hawking radiation \cite {GibHaw}, or the
horizon in de Sitter space. 

We will return to these and other arguments
at the end of this paper. For now, our main goal is to consider the
simplest of all inflationary scenarios, that of a massive field on a
true de Sitter background, and to study the conditions under which vacuum
fluctuations in this model evolve into classical fluctuations.

The essence of the method is given in Sec. II, and its specification
to the de Sitter case in Sec. III. Explicit examples for different values
of mass are discussed in Sec. IV. Section V is the summary of the main 
results and their implications, including the comparison with other works.

\section{Particle production}

Let $\phi$ denote a free, minimally coupled scalar field of mass 
$m$, and $S(\eta)$ the scale factor for the Robertson-Walker line element 
with  conformal time.
After the substitution $\phi \equiv \chi / S$, one can represent the amplitude
$\chi (\vec {x}, \eta)$ as the sum or the 
integral of the modes $\chi_k(\eta ) Q_{\vec {k}}(\vec {x})$, so that
$\nabla^2 Q_{\vec {k}} = - k^2 Q_{\vec {k}}$. The explicit form of the 
spatial modes depends on the geometry of the three-dimensional space. In either
case the field amplitudes are those for a set of harmonic oscillators
with time dependent frequencies:

\begin{equation}
\chi_k^{\prime \prime} + \omega_k^2 (\eta) \chi_k = 0~~,
\label {osceq}
\end{equation}
with
\begin{equation}
\omega_k^2(\eta) \equiv m^2 S^2 + k^2 - \frac {S^{\prime \prime}}{S}~~.
\label {omega}
\end{equation}

\noindent
Since, for a free field, the modes decouple and each of them may be considered
separately, from now on we can drop the subscript $k$. 

If $p$ denotes the canonical momentum of a mode $\chi$
we may define the one-mode creation and annihilation
operators $a$ and $a^{\dagger}$: 

\begin{equation}
a = (-)i~ \sqrt { \frac { |\omega(\eta)|}{2}}~\chi +
\frac {1}{\sqrt {2|\omega (\eta)|}}~ p ~~,
\end{equation}
\begin{equation}
a^{\dagger} = (+)i~
\sqrt {\frac {|\omega(\eta)|}{2}}~ \chi +
\frac {1}{\sqrt {2 |\omega(\eta)|}}~ p~~.
\end{equation}

These definitions agree with the standard ones when the frequency is real,
and the potential is upright. This assures that we will have the same 
initial vacuum states and particle excitations as in the textbook
cases. When the frequency is imaginary, and the potential upside-down, 
the presence of $|\omega|$ rather than
$\omega$ assures that $a$ and $a^{\dagger}$ obey the 
creation/annihilation algebra, and that the particle number still makes sense in
that regime. One should be aware though that when the frequency is imaginary, 
the Hamiltonian of the excitations is nothing like $\hbar \omega$.
This is the price that we have to pay for having the sensible definition
of creation and annihilation operators for an upside-down oscillator, and 
a sign of the very different physics in this case.

We can now define the one-mode number operator $n \equiv a^{\dagger} a$.
Its rate of change, for our case, is found to be

\begin{equation}
\frac {dn}{d\eta} = \frac {i}{2} 
\frac {\omega^2 - |\omega|^2}{ |\omega|} (a^{\dagger 2} - a^2) +
\frac {1}{2} \frac {|\omega|^{\prime}}{|\omega|} ( a^{\dagger 2} + a^2)
~~.
\label {dn}
\end{equation} 

The two terms in Eq. (\ref {dn}) describe the two sources of particle production:
the possible upside-down shape of the potential, and the possible change in
curvature of the potential, either upside-down or upright, with time.
In particular, if the initial state is the Fock vacuum with  $n =0$, a finite 
value may develop in time.

In most of the cases that have been considered before \cite {partprod}
only the second term contributes. In some of them the rate of change of 
frequency is small, and the particle number may still be defined as an 
adiabatic invariant. 
However, if the frequency is imaginary, 
there will be some particle production due to the first term, and the
adiabatic invariance of the particle number will not apply. It will be seen
in a moment how this works in the de Sitter case.

Although Eq. (\ref {dn}) provides a good illustration of what is going
on, it does not offer an easy way to calculate the time evolution
of the number operator. For this I will use the method of Ref.\cite {method}, 
with a slight generalization to cover the case of the upside-down potential.

Let $\chi_i$ and $\chi_i^{\prime}$ be the initial conditions at some
$\eta_i$, 
and $\chi_1(\eta)$ and $\chi_2(\eta)$ are the two independent solutions
of the oscillator equation (\ref {osceq}). The general solution may
be written as

\begin{equation}
\chi(\eta) = A(\eta) \chi_i + B(\eta) \chi_i^{\prime}~~,
\end{equation}
\noindent
with time dependent coefficients
\begin{equation}
A(\eta) \equiv \frac {\chi_{1}(\eta) \chi_2^{\prime}(\eta_i) - 
\chi_{1}^{\prime}(\eta_i) \chi_{2}(\eta)}
{W(\eta_i)} ~~,
\label {A}
\end{equation}
\begin{equation}
B(\eta) \equiv \frac {\chi_{1}(\eta_i) \chi_{2}(\eta) - 
\chi_{1}(\eta) \chi_{2}(\eta_i)}
{W(\eta_i))} ~~,
\label {B}
\end{equation}
\begin{equation}
W(\eta_i) \equiv 
{\chi_{1}(\eta_i) \chi_{2}^{\prime}(\eta_i) - 
\chi_{1}^{\prime}(\eta_i) \chi_{2}(\eta_i)} ~~.
\end{equation}

\noindent
Using  $A$ and $B$  one can calculate the off-diagonal 
Bogoliubov coefficient $b_{12}$:

\begin{eqnarray}
b_{12}(\eta) = (-)\frac {i}{2} A(\eta)
\left | \frac {\omega(\eta)}{\omega(\eta_i)} \right |^{1/2} &+&
\frac {1}{2} B(\eta) 
\left | \omega (\eta) \omega (\eta_i) \right |^{1/2} \nonumber \\  &+&
\frac {1}{2}
\frac {A^{\prime}(\eta)}{\left | \omega(\eta) \omega(\eta_i) \right |^{1/2}}
+
\frac {i}{2} B^{\prime}(\eta)
\left | \frac {\omega(\eta_i)}{\omega(\eta)} \right | ^{1/2}.
\label {off}
\end{eqnarray}

\noindent
The one-mode occupation number is then given as

\begin{equation}
\langle 0 | n (\eta) |0 \rangle 
= |b_{12} (\eta) |^2 ~~,
\label {n}
\end{equation}
\noindent
where $| 0 \rangle$ refers to the Fock vacuum at the initial time $\eta_i$.
This is the main quantity we are interested in. If the occupation number is
large, we have that 
$\langle a a^{\dagger} \rangle = 
\langle a^{\dagger} a \rangle [ 1 + {\cal O} (1/n)]$, so, to a high accuracy
 $a$ and $a^{\dagger}$ commute, $\chi$ and $p$ also approximately
commute, and the behavior of the mode is essentially classical. 
This elementary criteria for classicality has been used before in the
Schroedinger picture by Guth and Pi \cite {GuthPi}. Another Schroedinger
picture criteria, developed by Lyth \cite {Lyth}, will be discussed in the 
Appendix. 

Note that the canonical commutator has the same form for any choice of creation 
and annihilation operators, so the above criteria for classicality is immune 
to the usual ambiguity of defining particles in curved spacetime.  The
particle production in this case leads to the formation of a classical
condensate, and quantum fluctuations are effectively decohered.  We would like 
to know what the conditions for this to happen in de Sitter space are
so that such decohered quantum fluctuations may serve as the source of 
classical density perturbations in inflationary cosmology.

\section{$\lowercase{k}=0$  De Sitter expansion}

For spatially flat De Sitter expansion
the scale factor evolves as $S = -(H_0 \eta)^{-1}$, with $\eta \in (- \infty,
0)$, and $H_0 = {\rm const}$. Introducing a new independent variable
$z \equiv - k\eta$, the one-mode frequency (\ref {omega}) becomes

\begin{equation}
\omega^2 (z) = 1 + \frac {1}{z^2} \left ( \frac {m^2}{H_0^2} - 2 \right )~~.
\label {omegads}
\end{equation}

The modes themselves are well known:

\begin{equation}
\chi_1(z) = \frac {\sqrt {\pi}}{2} \sqrt {z} H_{\nu}^{(2)}(z)~~,~~
\chi_2(z) = \frac {\sqrt {\pi}}{2} \sqrt {z} H_{\nu}^{(1)}(z)~~.
\end{equation}
$H^{(1,2)}$ are Hankel functions, and $\nu \equiv \sqrt {9/4 - m^2/H^2}$. 
These modes are normalized so that $W (\eta) = i$ at all times. The
minus sign in the definition of variable $z$ is to avoid complications
due to the branch cut of Hankel functions on the negative real axis. Thus,
the flow of time is from large to small values of $z$.
At early times, when $z_i \gg 1$, the modes reduce to plane waves

\begin{equation}
|\omega(z_i)| = 1~~,~~
\chi_{1,2}(z) = \frac {1}{\sqrt {2}} e^{\mp i(z - \nu \pi/2 - \pi/4)}~~.
\label {initial}
\end{equation}

The expression for the off-diagonal Bogoliubov coefficient is now
simplified,

\begin{equation}
b_{12}(z) = \frac {1}{2} |\omega(z)|^{1/2} [ B(z) - i A(z) ]
+ \frac {1}{2 |\omega|^{1/2}} [A^{\prime}(z) + i B^{\prime}(z) ]~~,
\end{equation}
\noindent
and a straightforward calculation leads to our final expression for
the number of particles produced in course of the expansion

\begin{equation}
n(z) = \frac {1}{2|\omega(z)|}|\chi_2^{\prime}(z)|^2 +
\frac {|\omega(z)|}{2} |\chi_2(z)|^2 - \frac {1}{2}~~.
\label {nnu}
\end{equation}

\noindent
This expression does not depend on the initial
moment $z_i$, but it correctly reproduces the initial vacuum condition
$n (\infty) = 0$ in the limit $z \rightarrow z_i \rightarrow  \infty$.
 
However, it is the opposite regime, $z \rightarrow 0$, the regime of
large expansion, that is of most interest, as at that time the wavelength
of the mode exceeds the Hubble radius.
This is one way that inflation solves the horizon 
problem. 

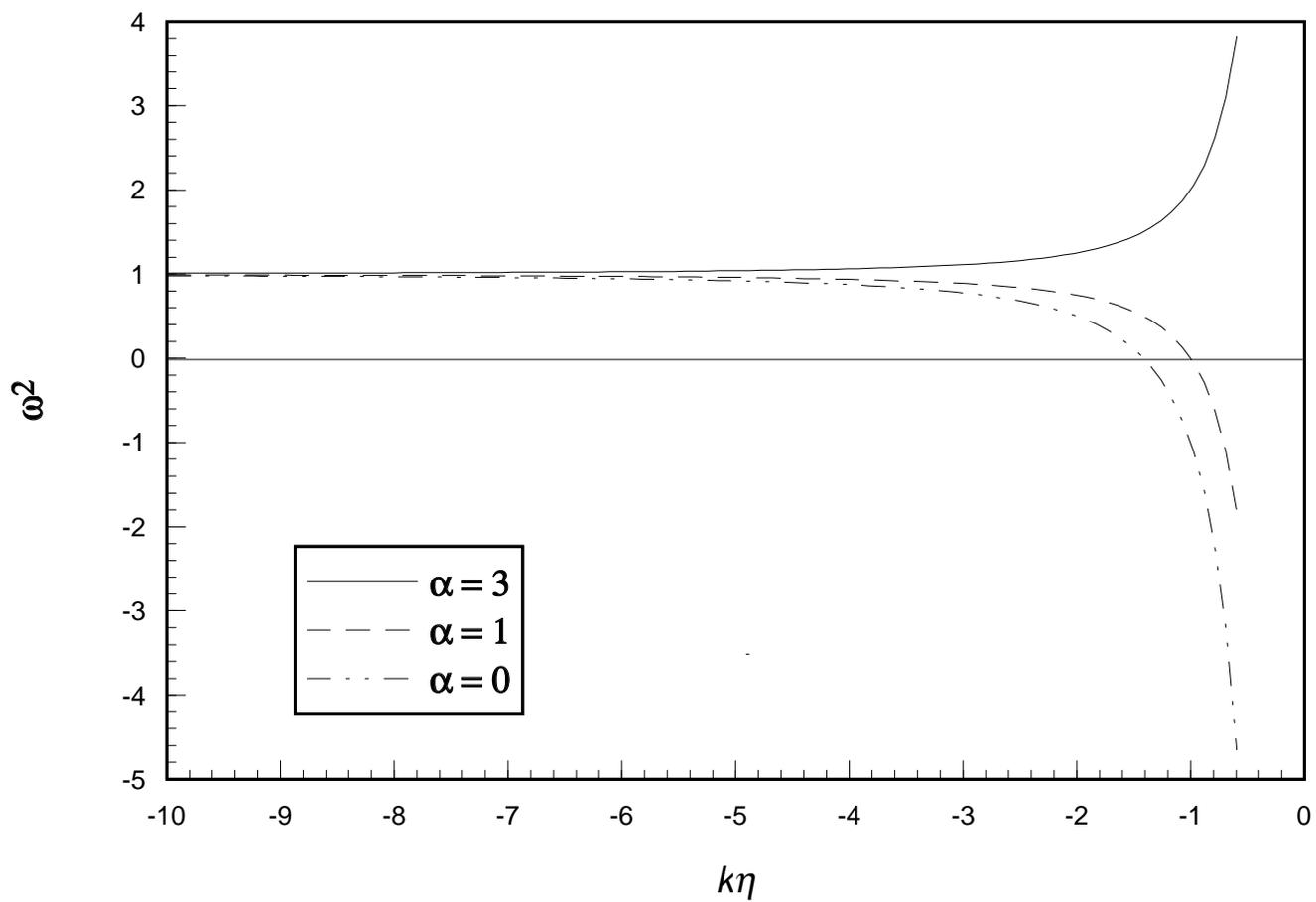
\begin{figure}
\caption{The evolution of the one-mode frequency squared in de Sitter
space, Eq. (\ref {omegads}), for different values of mass parameter 
$\alpha \equiv m^2/H_0^2$.
In all cases the frequency starts to change rapidly after the physical
wavelength of the mode exceeds the Hubble radius. In the same regime the
one-mode quantum oscillators turn to an upside-down shape for $\alpha < 2$.}
\label{fig1}
\end{figure}

The behavior of frequency Eq. (\ref {omegads}) in this regime may be
seen from Fig.\ \ref{fig1}.  There are
two cases: real frequency and bounded classical motion for 
$m^2 > 2 H_0^2$, and imaginary frequency and \lq\lq rolling down the hill
\rq\rq$~$ for $m^2 < 2 H_0^2$. The first term in Eq. (\ref {dn})
affects the particle number at 
$z < 1$ only for $m^2 < 2 H_0^2$, and is absent for larger masses.

The moment when these effects turns on is about or after the moment at which
the fluctuation crosses the Hubble radius. The frequency vanishes at
$z_0 = \sqrt {2}~\sqrt {1 - m^2/(2 H_0^2)}$. Compared to the moment
of crossing the Hubble radius, the two scale factors are related as

\begin{equation}
S(\eta_{0}) = S (\eta_{c}) \left ( 2 - \frac {m^2}{H_0^2} \right )^{-1/2} ~~.
\end{equation}

The scalar field which drives inflation is supposed to have
a very low mass,  $m^2 \ll H_0^2$, and in that case, and up to a mass as
large as $H_0$, the one-mode frequency turns imaginary just before the 
mode crosses the Hubble radius. However, for $H_0^2 < m^2 < 2 H_0^2$, 
the \lq\lq rolling down the hill\rq\rq $~$is postponed to later times, 
well after the physical wavelength of the mode equals the Hubble radius. 
Either way, it is the evolution at $z <1$ that decides the fate of
the fluctuation.

The other source of change in the particle number is the variation of frequency
with time. This is given as

\begin{equation}
\frac {|\omega|^{\prime}}{|\omega|} = 
\frac {m^2/H_0^2 -2}{|z| ( m^2/H_0^2 -2 + z^2)}~~.
\label {omegarate}
\end{equation}

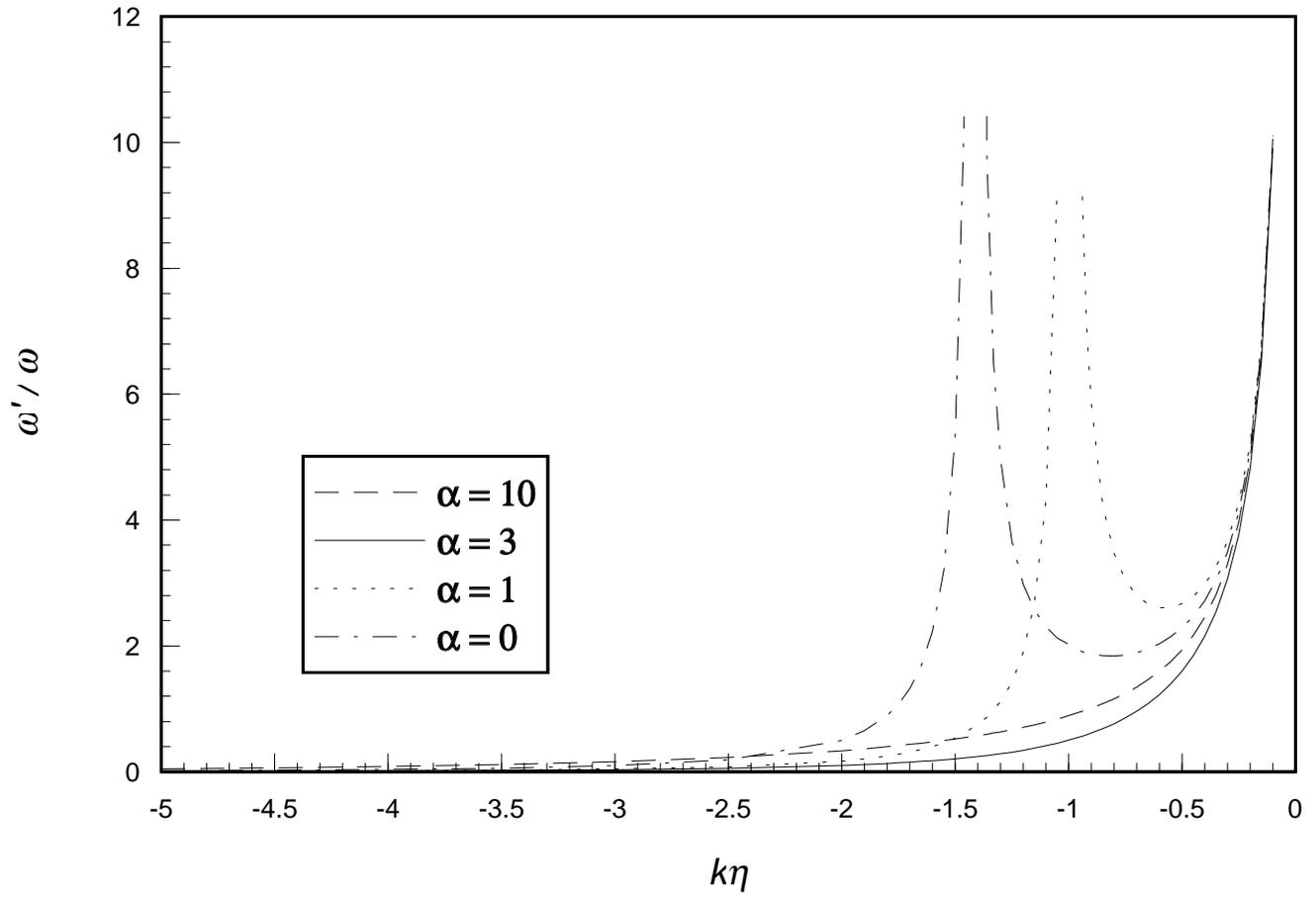
\begin{figure}      
\caption{The evolution of a relative change of the one-mode frequency
in de Sitter space, Eq. (\ref {omegads}), for different values of mass.
For $\alpha < 2$ the frequency passes through zero, and there is a 
spurious singularity in $|\omega^{\prime}/\omega|$. For all values of mass
the relative change in frequency exceeds unity shortly after the wavelength
surpasses the Hubble radius.}
\label{fig2}
\end{figure}

Figure\ \ref{fig2} shows the behavior of this function for several 
characteristic values of $m^2/H_0^2$. The universal growth as 
$z\rightarrow 0$ is due to the accelerating expansion of the de Sitter universe.
For all values of mass the rate of change of frequency exceeds unity very 
soon after the mode crosses the Hubble radius. Thus, the second term
in Eq. (\ref {dn}) gives the major contribution to the change in particle
number at $z <1$ for all values of mass. 

There is another regime with a large frequency change in the 
$m^2 < 2 H_0^2$ case,
near point  $z=z_0$  where the frequency vanishes. This period
lasts just an e-fold of expansion or so, before the rate of change
in frequency drops close to unity and then starts to grow with  
$z \rightarrow 0$. In this case we have a shrinking size of the phase space
cell, rather than its large occupancy, so the expected spike in particle
number should be considered as a spurious effect.

To proceed, it will be both sufficient and convenient to evaluate Eq. 
(\ref {nnu}) in the approximation of small $z$. Consider first the small
argument expansion for Hankel functions when $\nu$ is not an integer
\cite {GrRy}. For the $m^2 > 0$ cases this means $\nu \not= 0, 1$.
The relevant mode and its derivative may be expressed as

\begin{equation}
\chi_2(z) = A_{\nu}~z^{\nu + 1/2} + 
B_{\nu}~z^{1/2 -\nu}
+{\cal O}(z^{\nu + 5/2}, z^{- \nu + 5/2}) ~~,
\end{equation}
\begin{equation}
\chi_2^{\prime}(z) = (\nu + 1/2)A_{\nu}~z^{\nu - 1/2} 
+ (- \nu + 1/2)B_{\nu}~z^{-\nu - 1/2}~~.
\end{equation}
\noindent
The  $\nu$-dependent coefficients are defined as

\begin{equation}
A_{\nu} \equiv \frac {\pi^{1/2}e^{i\pi\nu}}{2^{\nu +1} \Gamma (\nu + 1)}
\left ( 1 + i \cot (\nu \pi) \right )~~,~~
B_{\nu} \equiv \frac {ie^{i\pi \nu}}{\pi^{1/2}} 2^{\nu -1} \Gamma (\nu)~~.
\end{equation}
\noindent
The $e^{i\pi \nu}$ factors assure that these modes maintain $W =i$
normalization. In two special cases the small argument expansions of the
modes have logarithmic corrections (which do not affect the qualitative
behavior of the solutions):

$\nu =0$,
\begin{equation}
\chi_2(z) = A_0 z^{1/2} ~+~ B_0 z^{1/2} \ln \frac {z}{2}~
+~{\cal O}(z^3) ~~,
\end{equation}
\begin{equation}
\chi_2^{\prime}(z) =\left ( \frac {A_0}{2} + B_0 \right ) z^{- 1/2} 
~+~ \frac {B_0}{2} z^{- 1/2} \ln \frac {z}{2}~~,
\end{equation}
\noindent
with
\begin{equation}
A_0 \equiv  \frac {\pi^{1/2}}{2} + i \frac {C}{\pi^{1/2}}~,~~
B_0 \equiv \frac {i}{\pi^{1/2}}~~.
\end{equation}

\noindent
$\nu = 1$
\begin{equation}
\chi_2(z) = A_1 z^{3/2} ~+~ B_1 z^{3/2}  \ln \frac {z}{2}~
+~ C_1  z^{-1/2}~
+{\cal O}(z^2) ~~,
\end{equation}
\begin{equation}
\chi_2^{\prime}(z) = \left ( \frac {3}{2}A_1 + B_1 \right )z^{1/2} 
~+~ \frac {3}{2}B_1 z^{1/2} \ln \frac {z}{2}~-~
\frac {C_1}{2} z^{-3/2}~~.
\end{equation}
\begin{equation}
A_1 \equiv \frac {\pi^{1/2}}{4} + 
i \frac {2C -1}{4\pi^{1/2}} ~,~~
B_1 \equiv  \frac {i}{2\pi^{1/2}} ~,~~
C_1 \equiv - \frac {i}{\pi^{1/2}}~~.
\end{equation}

\noindent
$C$ is the Euler constant.

The behavior of the modes at small $z$ is easy enough to understand. 
For non-integer $\nu$ there are three different regimes: (a)  $ m^2/H^2 < 2$ 
or $ \nu > 1/2$, (b) $ 2 < m^2/H_0^2 \leq 9/4$ or $1/2 > \nu \geq  0$, (c) $m^2/H_0^2 > 9/4$ or $\nu \in i{\cal R}$.

In the first case the potential turns upside-down at $z_0$ and the field 
amplitude increases without bound, $\chi_2 \sim |z|^{1/2-\nu}$ as 
$z \rightarrow 0$. In the second case, the potential is always upright 
but becomes rather shallow at $z= 1$, below which the small argument
expansion starts to apply. The $z^{1/2-\nu}$ mode dominates again, 
only now it describes slow settling to the minimum. Finally, for $\nu$ 
imaginary we have an oscillatory approach to a
minimum at a rate proportional to $z^{1/2}$. The relative strength of 
the two modes depends on the mass. In all cases the derivative diverges,
as the potential is getting steeper and narrower around $\chi = 0$.

This elementary discussion will be sufficient to understand what happens
to the particle number. Using the preceding small argument expansion,
and noting that the frequency may be approximated in the same regime as
$|\omega| \rightarrow |1/4 - \nu^2|^{1/2} z^{-1}$, Eq. (\ref {nnu}) becomes

\begin{eqnarray}
n(z) = &-& \frac {1}{2}~ \nonumber \\
&+& \frac {1}{2} \left | \frac {1}{4} - \nu^2 \right |^{1/2}
\left  [ 
|A_{\nu}|^2 z^{\nu + \nu^*} +
|B_{\nu}|^2 z^{-(\nu + \nu^*)}
+ A_{\nu} B_{\nu}^{*} z^{\nu - \nu^*}
+ A_{\nu}^{*} B_{\nu} z^{\nu^* - \nu}\right ] ~~
\nonumber \\
&+& \frac {1}{2 |1/4 - \nu^2 |^{1/2}}
[
|\nu + 1/2|^2 |A_{\nu}|^2 z^{\nu + \nu^*} +
|- \nu + 1/2|^2 |B_{\nu}|^2 z^{-(\nu + \nu^*)} 
\nonumber \\
&+& (\nu + 1/2)(-\nu^* +1/2) A_{\nu}^{\prime} 
B_{\nu}^{*} z^{\nu - \nu^*} ~ \nonumber \\
&+&(-\nu + 1/2)(\nu^* +1/2) A_{\nu}^{*} 
B_{\nu} z^{\nu^* - \nu}  ]~.
\label {nsmall}
\end{eqnarray}

In the two special cases when $m^2 >0$ and $\nu$ is an integer, 
we have, at $z \ll 1$,

\begin{equation}
n(z) = \frac {1}{\pi \sqrt {3}} \frac {1}{z^2} ~
~~~~~
{\rm if}~~ \nu =1~~;
\label {nu1}
\end{equation}
\begin{equation}
n(z) = \frac {1}{2\pi} \left | \ln \frac {z}{2} \right |^2
~~~~~~ {\rm if}~~ \nu =0 ~~.
\label {nu0}
\end{equation}
We can now see how the mass of the scalar field
affects the amount of particle production.

\section{The effect of mass}

Since from very early on it has been realized that the inflaton field
needs to be very weekly (self-)coupled, or nearly massless 
\cite {fluctuations} most of the work on fluctuations in inflationary
cosmology concentrated on those cases. 
There are very few results about the behavior of the fields
with mass of the order of the Hubble parameter, and almost none for
heavier fields. Yet, there have been clear indications that the behavior of 
massive scalars in De Sitter space very much depends on the mass. Equation
(\ref {nsmall}) brings this expectation to the full force, and immediately
yields our main result: The particle number for heavy fields in De Sitter space,
with $m^2/H_0^2 > 9/4$, is an oscillatory function at times 
later than the Hubble radius crossing time. There is some particle 
production, but there are no classical condensates with $n \gg 1$. In contrast,
the particle number for medium and light fields in De Sitter
space, with $0\leq m^2/H_0^2 < 9/4$, 
except $m^2 = 2 H_0^2$, diverges at times later than the
Hubble radius crossing time. There is a significant particle production,
leading to the formation of classical condensates with $n \gg 1$.

The  $m^2 = 2 H_0^2$ case is
conformally related to flat spacetime. The frequency vanishes for
this choice of mass, the modes are plane waves at all times, and there
is no particle production. This is one version of a classic result by
Parker \cite {partprod}.

Let us now illustrate these results by taking a closer look at the specific
cases.

(a) $m^2/H_0^2 = - 4$, $\nu = 5/2$. This kind of model was analyzed by Guth and Pi \cite {GuthPi} as 
it has direct bearing for the generation of density perturbations in the new 
inflationary scenario. The $|m^2|/H_0^2$ ratio is supposed to be very small,
but for illustrative purposes I will take it here to be large, so that
the parameter $\nu$ is a half-integer, and the modes are given through
elementary functions. We can then calculate the particle number at all times. 
This is not a limitation, since,
according to Eq. (\ref {nsmall}), the behavior of the particle number is
qualitatively similar for all $\nu > 0$.
We find

\begin{equation}
\chi_2 (z) = \frac {1}{\sqrt{2}} \left [ - \frac {3}{z} + i \left (
1 - \frac {3}{z^2} \right ) \right ] ~e^{iz}~~,~~
\omega^2(z) = 1 - \frac {6}{z^2}~~
\end{equation}
and

\begin{equation}
n(z) = \frac {z - 3/z^2 + 36/z^6}{4 |z^2 -6|^{1/2}} +
\frac {1}{4} |z^2 - 6|^{1/2} \left ( \frac {1}{z} + \frac {3}{z^3} +
\frac {9}{z^5} \right ) - \frac {1}{2} ~~.
\label {CW}
\end{equation}

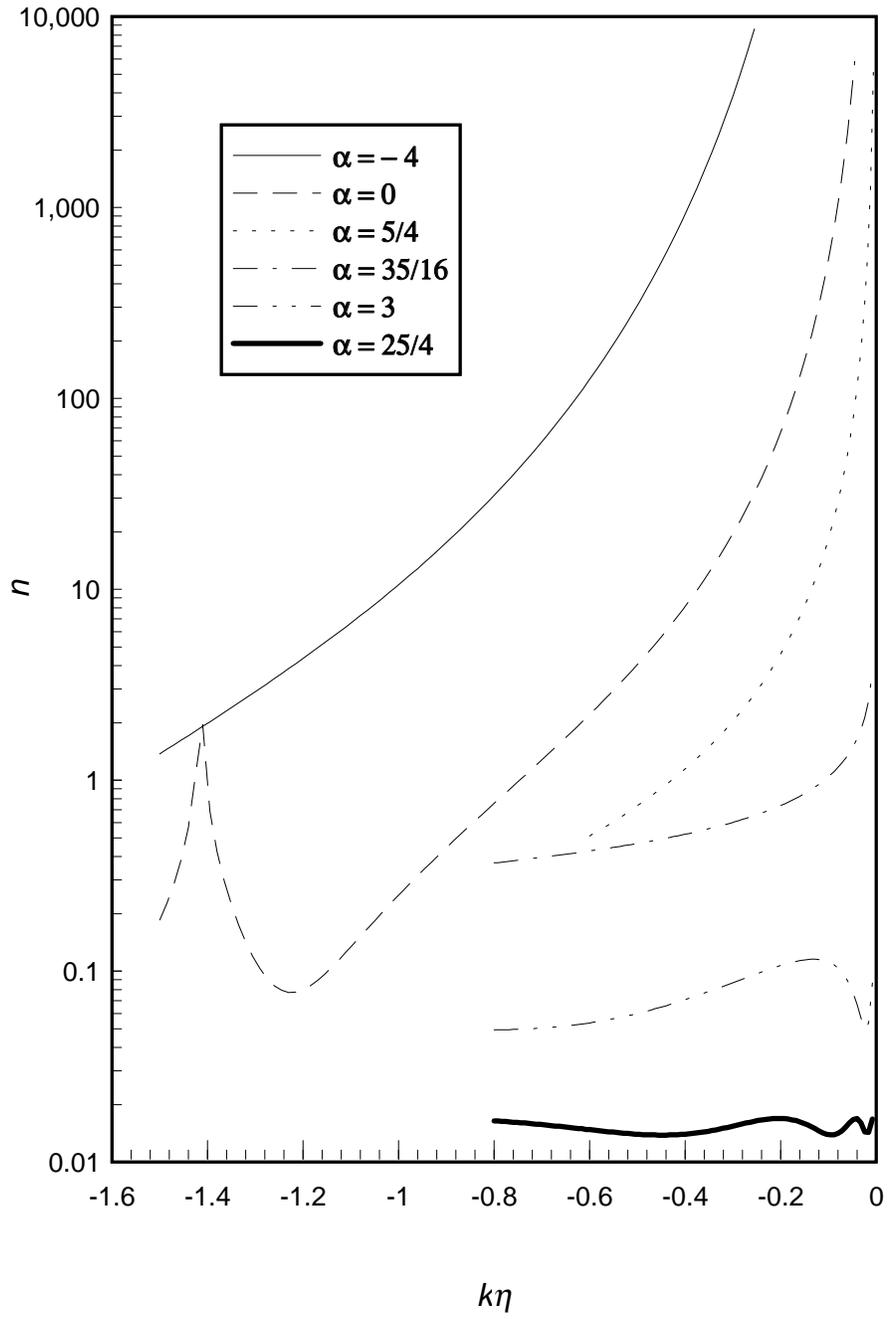
\begin{figure}
\caption{The evolution of a one-mode occupation number in De Sitter
space for different values of mass, the main result of this work. 
The \lq\lq Coleman-Weinberg \rq\rq $~$ type case with a large mass, 
$|\alpha|=4$,
Eq. (\ref {CW}), and the fiducial $m^2 =0$ case, Eq. (\ref {nformeq0}), 
are plotted for the whole range, others only for $ |k \eta| \ll 1$, when the
mode is well outside the Hubble radius. The divergence at 
$k\eta =-2^{1/2}$ for $\alpha =0$ is a spurious effect, due to $\omega^2$
passing through zero as the oscillator turns into an upside-down shape.
The corresponding divergence in the $\alpha = -4$ case is just outside the range
of the figure. The $\alpha =5/4$ case corresponds to $\nu =1$, and, 
as a special case, it is given by Eq. (\ref {nu1}). In $\alpha = 35/16$, 
or $\nu = 1/4$ case,
the particle number may be evaluated from Eq. (\ref {upside}) as
$n(z) \approx 0.33/z^{1/2}$. Unlike the preceding two cases the quantum
oscillator has upside-right shape at $|k \eta |\ll 1$, but the particle
number nevertheless diverges due to the rapid change in frequency. 
The two cases with imaginary $\nu$ are $\alpha = 3$ or $\nu = i 3^{1/2}/2$,
and $\alpha = 25/4$ or $\nu = 2 i$. Both are given through Eq. (\ref {nosc}).
The particle number is bounded, oscillatory function of $\ln |k \eta|$,
and there is no formation of classical condensates.}
\label{fig3}
\end{figure}

The behavior of this function is shown in Fig.\ \ref{fig3}. The oscillator turns
upside down while the mode is still within the Hubble radius, and soon after 
the particle number rapidly increases,
approaching quickly a $z^{-2\nu}$ growth for wavelength longer than the
Hubble radius.

(b) $m^2 =0$, $\nu = 3/2$. This model is the starting point for most inflationary scenarios. The modes are again  given through elementary functions

\begin{equation}
\chi_2 (z) = \frac {e^{-i\pi}}{\sqrt {2}} \left ( 1 + \frac {i}{z}
\right ) e^{iz} ~~,~~
\omega^2(z) = 1 - \frac {2}{z^2} ~~,
\label {masslesschi}
\end{equation}

\noindent
and the particle number Eq. (\ref {nnu}) may be computed 
through all times, not just at small $z$. The result is 

\begin{eqnarray}
n(z) = (-) \frac {1}{2} &+& \frac {1}{4} 
|z^2 -2|^{1/2} \left ( \frac {1}{z} + 
\frac {1}{z^3} \right ) ~ \nonumber \\
&+&
\frac {1}{4 |z^2 -2|^{1/2}} \left ( z - \frac {1}{z}
+ \frac {1}{z^3} \right )
~~.
\label {nformeq0} 
\end{eqnarray}

The divergence at $z_0 = \sqrt {2}$, is due to the vanishing frequency
at this point. It is represented by the spike on Fig.\ \ref{fig3}.
$n$ quickly returns 
to value below unity as soon as the mode crosses the Hubble radius, and 
then the rapid particle production starts in earnest. 
For long-wavelength modes, well outside 
the Hubble radius ($z \ll 1$), we have that the occupation number 
diverges as $z^{-3}$. This may be written as

\begin{equation}
n(\eta) \sim \frac {3}{4 \sqrt{2}} \left ( \frac {\lambda_{phys}
(\eta)}{H_0^{-1}} \right ) ^3 ~~.
\end{equation}

(c) $0 \leq  m^2/H_0^2 < 2$, $\nu \in (1/2, 3/2]$. The result for $\nu =1$ is given before Eq. (\ref {nu1}). For $\nu \not= 1$ the expression for particle number at small $z$ becomes

\begin{equation}
n(z) = \nu \left [ \frac {\nu - 1/2}{\nu + 1/2} \right ]^{1/2}~
\frac {2^{2\nu - 2}}{\pi} |\Gamma (\nu)|^2~ z^{-2 \nu} ~~.
\end{equation}

Both the growth rate and the overall normalization of this function
decrease with the mass square increasing from $0$ to a zero production
value of $2 H_0^2$. The behavior smoothly changes as $\nu$ decreases,
and it is well represented by the $\nu =1$ case in Fig.\ \ref{fig3}.

(d) $2 < m^2/H_0^2 \leq 9/4$, $\nu \in [0, 1/2)$. Unlike the preceding case, 
the frequency now stays real and the potential is upright at all times, including times later than the Hubble crossing
time. Nevertheless, as Eq. (\ref {nsmall}) shows, the particle number 
diverges as $z \rightarrow 0$:

\begin{equation}
n(z) = \left [ \frac {1/2 - \nu}{1/2 + \nu} \right ]^{1/2}
~\frac {2^{2 \nu -3}}{\pi} |\Gamma (\nu)|^2~z^{- 2\nu} ~~.
\label {upside}
\end{equation}

One example of this kind, with $m^2/H_0^2 = 35/16$ and $\nu = 1/4$, is
illustrated on Fig.\ \ref{fig3}. The behavior varies smoothly with the increasing
mass, until  $\nu =0$, given separately in (\ref {nu0}), when
the growth drops from a power law to the logarithm.
Since the frequency is real, the particle production in all of these cases 
must be entirely due to the rapid change of frequency at
small $z$. The relative shallowness of the potential early on (just after
the Hubble radius crossing) probably plays some role as well.

(e) $m^2/H_0^2 > 9/4$, $\nu \in i {\cal R}$. In this case the behavior of 
$n(z)$ at $z \ll 1$ is purely oscillatory. An
explicit substitution of $\nu = i |\nu|$ to Eq. (\ref {nsmall}) gives the
following result:

\begin{equation}
n(z) = A_0 ~ +~
 A_c \cos \left [ 2 |\nu| \ln (z/2) - 2 \Phi_{\Gamma} \right ]
~+~
A_s \sin \left [ 2 |\nu| \ln (z/2) - 2 \Phi_{\Gamma} \right ] ~~.
\label {nosc}
\end{equation}

\noindent
$\Phi_{\Gamma}$ is the phase of $\Gamma (i|\nu|)$ \cite {GrRy}.
The three coefficients are found to be

\begin{equation}
A_0 \equiv (-) \frac {1}{2} + \frac {1}{2}~
\frac {( 1/4 + |\nu|^2 )^{1/2}}{|\nu|} ~
\coth (|\nu|\pi) ~~,
\end{equation}
\begin{equation}
A_c \equiv 
\frac {3/4 - |\nu|^2}{2|\nu|}~  
 \frac {1 + \coth (|\nu|\pi)}{4(1/4 + |\nu|^2 )^{1/2}}~ 
e^{- \pi |\nu|}~~,
\end{equation}
\begin{equation}
A_s \equiv (-)~
 \frac {1 + \coth (|\nu|\pi)}{4(1/4 + |\nu|^2 )^{1/2}}~ 
e^{- \pi |\nu|}~~.
\end{equation}

The period of the oscillations is $z_1/z_2 = \exp [\pi/|\nu|]$. 
Figure \ \ref{fig3} shows two such cases. As in the case (d),
the potential is upright, and the only source of particle production 
is the rapid variation in frequency. However, unlike the previous case,
the steeper potential does not allow for a large amplitude in $\chi$,
and the change in frequency does not lead to a large particle number,
merely to oscillations in its value. The particle number is bounded by,
$n(z) \leq f \equiv A_0 + ( A_c + A_s)^{1/2}$. If we denote $x = |\nu|$, this
amplitude may be expressed as

\begin{equation}
f(x) = - \frac {1}{2} + \frac {(1/4 + x^2)^{1/2}}{2 x} \coth (\pi x) +
\frac {R(x)^{1/2}}{8 \sinh (\pi x)} ~~, x \in (1, \infty)~~,
\label {bound}
\end{equation}
\noindent
where
\begin{equation}
R(x) \equiv \frac {x^4 + 5x^2/2 + 9/16}{x^2(x^2 + 1/4)}~~.
\end{equation}

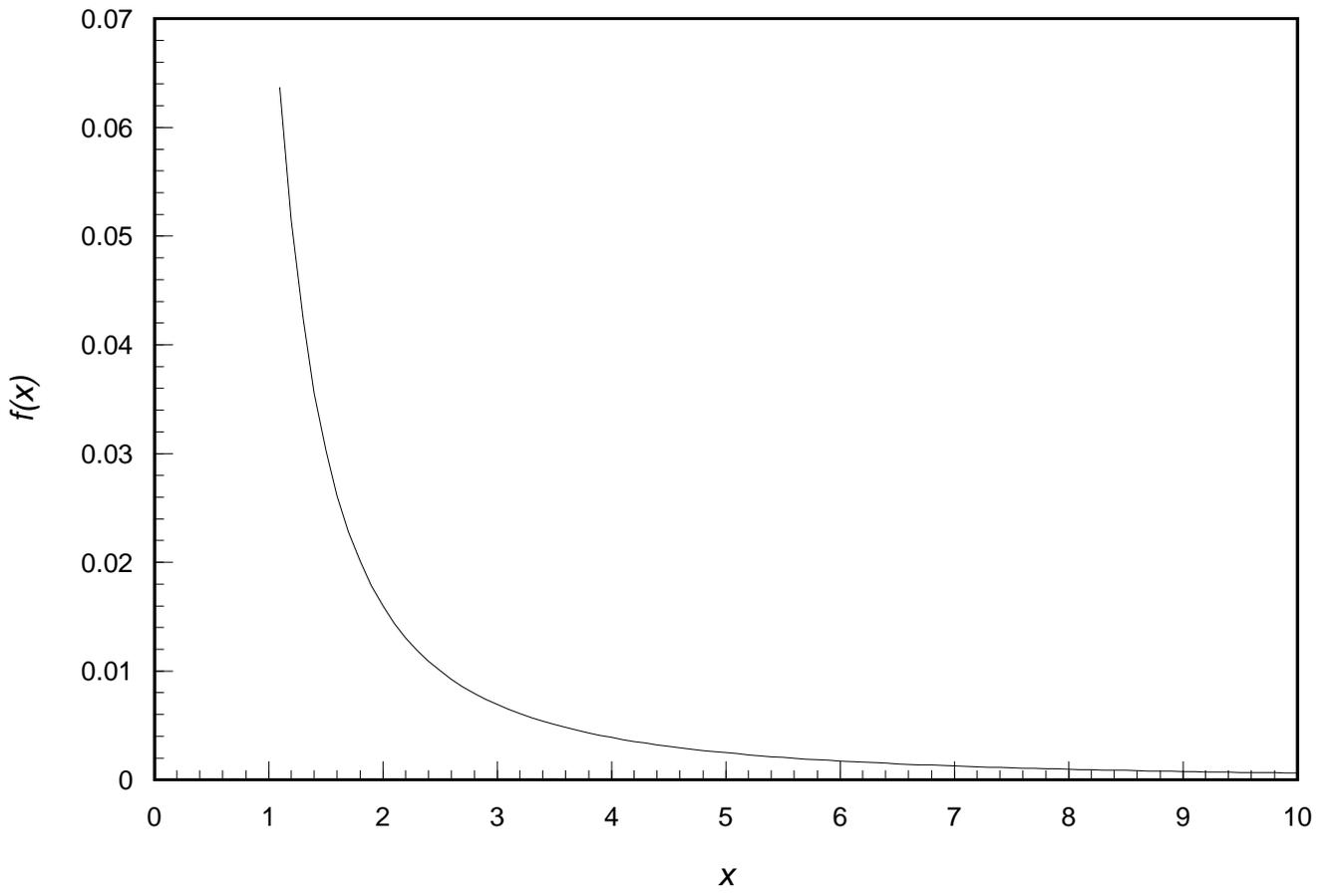
\begin{figure}
\caption{The maximal amplitude of the particle number for heavy fields, 
$m^2 > 9H_0^2/4$, Eq. (\ref {bound}), as function of mass, $x \equiv |\nu|$. 
This surprisingly smooth function shows that in
course of its oscillations the particle number never exceeds unity, so
the vacuum fluctuations of heavy fields in de Sitter space do not
have a classical behavior even for wavelengths that exceed the Hubble 
radius.}
\label{fig4}
\end{figure}

As Fig.\ \ref{fig4} shows,  this amplitude is a remarkably smooth function of $|\nu|$. 
The growth as $\nu \rightarrow 0$ is
due to the inaccuracy of the employed small $z$ expansion for noninteger $\nu$ 
near integer value $\nu =0$. The transition from the bounded and oscillatory
behavior of the mode at $\nu \in i {\cal R}$ to unbounded logarithmic 
growth when $\nu =0$ cannot be described accurately by the lowest order
expansion in $z$, and it is simulated as the divergence of $\Gamma (\nu)$
as $\nu \rightarrow 0$. However, for $|\nu| > 1$ the expansion may be trusted
and, as Fig.\ \ref{fig4} shows, the particle number is indeed bounded for all values
of mass. The amplitude of the oscillations tends to zero as $H_0^2/m^2$ in
the limit of very high mass. 

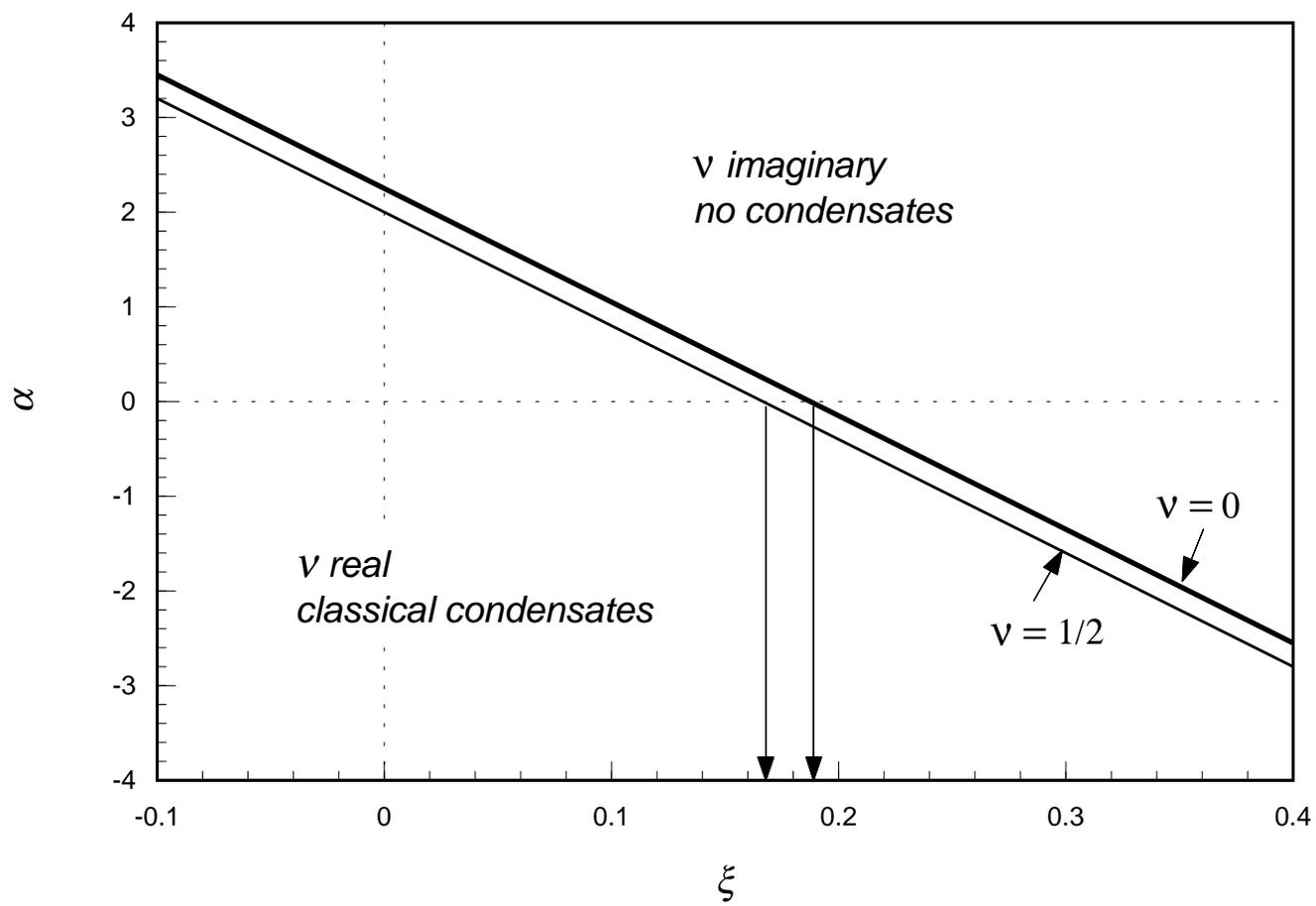
\begin{figure}
\caption{Formation of classical condensates and the decoherence of
vacuum fluctuations in De Sitter space as function of both
the mass and the nonminimal coupling. Condensates form not just for 
$\omega^2 < 1$ and $\nu \in {\cal R}$, but also for a narrow set of
values between $\nu =0$ and $\omega^2 = 1$ (or $\nu =1/2$). See part f,
Section 4.}
\label{fig5}
\end{figure}

(f) nonminimal coupling. The effect of adding an $\xi R \phi^2/2$ term is particularly simple in case of a $k=0$ de Sitter background: in all the expressions above one should
replace $m^2$ with $m^2 + 12 \xi H_0^2$. The behavior of the occupation
number as parametrized by different values of $\nu$ is the same as before,
except that now $\nu^2 \equiv 9/4 - m^2/H_0^2 - 12 \xi$, and classical
condensates form whenever this takes positive values or zero.
Figure \ \ref{fig5} shows the characteristic domains. The basic physics of the
phenomena is unchanged.

\section{Conclusions}

As we have seen, the ratio of the two scales, the Hubble radius $H_0^{-1}$ and
the Compton wavelength $m^{-1}$, determines the onset of two phenomena
in de Sitter space: the rapid rate of change in frequency of quantum 
oscillators and/or their turning into the upside-down shape. Together or 
separately these two phenomena lead to the formation of classical condensates 
and effectively classical behavior of quantum fluctuations.

In this picture parameter
$H_0^{-1}$ is important not because of its causal property, but because it is 
the characteristic scale for spacetime curvature. 
The decoherence scale varies from roughly $H_0^{-1}$ in the case of a massless 
field to infinity as mass of the field increases towards 
$m^2 = 2H_0^2$. The rapid growth in particle number is due to both the
upside-down shape of the quantum oscillators and the rapid change in their
frequency. For masses in the narrow range 
$m^2 \in (2H_0^2, 9H_0^2/4)$ the decoherence scale is again just a bit
larger than the Hubble radius. The potential is upright in this case,
and the formation of the condensate is entirely due to rapid expansion.
Finally, for fluctuations with even larger mass the particle number is
always bounded. These fluctuations do not decohere, and cannot serve as the
source of perturbations that seed galaxies.

The physical picture used here is just another side of the well known
phenomena of parametric amplification and squeezing of quantum
fluctuations \cite {LukNov} \cite {Grishchuk}, that played an important role
early on in the studies of particle production and density perturbations.
Two new moments covered here are the attention to decoherence through the formation of classical condensates, and the role of mass in the efficiency
of particle production and decoherence. 

It is a somewhat curious result that the $n(z) \sim z^{-2\nu}$ growth of the
occupation number for low and intermediate masses is the same as the
behavior of the dispersion ratio used by Guth and Pi \cite {GuthPi}
to characterize the late time classical behavior in similar models.
However, since they carried out their calculations for a model with a
potential of Coleman-Weinberg shape, the oscillators were upside-down
because of the curvature of that potential, and it was not clear from
Ref. \cite {GuthPi} whether the classical behavior would emerge if the
potential in the field Lagrangian was upright. Besides elucidating 
this role of mass, we obtained here a somewhat unexpected addendum to
Ref. \cite {GuthPi}: the rate of change of frequency may as well be just
as important for decoherence of low mass modes as the turning of the oscillators
to an upside-down shape. In fact, for intermediate mass fields, case (d) above,
this is the only source of decoherence.

Another curiosity has to do with the  kinematic argument that rapid expansion 
leads to the accumulation of modes with low, and almost indistinguishable 
frequencies. This \lq\lq infrared divergence\rq\rq$~$ supposedly
leads to a large occupation number for long-wavelength modes, hence to 
their highly classical behavior. This is a puzzling argument, since it
does not depend on the mass of the field, so one would expect the formation
of a classical condensate in all cases. The answer is that while the occupation
number in the massless case indeed looks as if it can be understood by a
kinematic argument $n \sim (\lambda_{phys}/H_0^{-1})^3$ for a finite
mass the exponent is smaller, and precludes kinematic interpretation.

The decoherence argument which is often considered to be the
most trustworthy is the demonstration of the effectively diagonal form of the 
reduced density matrix obtained by 
tracing out some  uninteresting degree(s) of freedom other than inflaton
field.  This general scheme appears to be the most satisfactory approach
to the classic \lq\lq measurement problem\rq\rq $~$ in quantum mechanics,
and its application to the creation of density perturbations in inflationary
cosmology has been investigated by several authors, e.g., Brandenberger {\it 
et al.} \cite {BLM}, Sakagami \cite {Sakagami}, and  Hu {\it et al.} \cite {HuPZ}. A common conclusion to all these approaches is
that the decoherence is proportional to the strength of the coupling, 
and it is supposed to disappear if the coupling goes to zero.  The analysis
of free models done here shows that in the de Sitter background there will be some decoherence which does
not depend on the magnitude of the coupling constant. The interaction or 
self-interaction may be sufficient but not a necessary condition for the 
decoherence. A similar conclusion may be reached by utilizing the stochastic
approach of Starobinsky \cite {StarMeudon} for the analysis of 
long-wavelength quantum fluctuations in an inflationary universe \cite {Nambu},
\cite {sdcgqf}.

\appendix

\section*{Comparison with Lyth's criteria for decoherence}

A plausible physical argument for the decoherence of quantum fluctuations
in an inflationary universe was pointed out by Lyth. \cite {Lyth} He
considered a massless, minimally coupled scalar field, initially in the
state of minimal uncertainty. Subsequently, the wave function of this 
time-dependent oscillator will spread. By treating this spread approximately
as in the case of a free particle in flat spacetime, Lyth was able to show
that times taken for the width to reach values that are large compared to the
initial spread are themselves quite large compared to the Hubble time. Due
to technical restrictions the argument applies only for a period of a few
Hubble times after the fluctuation crosses the Hubble radius, but, as Lyth
remarked, \lq\lq it seems plausible [...] that if the wave packet is already
extremely classical when this stage is reached, it will remain so for 60
or so Hubble times that the field survives.\rq\rq

Indeed, this argument is quite robust, and for a massless case it is 
equivalent to the criteria used here. To see this, recall that Lyth uses
the field amplitude before rescalling $\phi = \chi_2/S$. In the massless case,
$\chi_2$ is given by Eq. (\ref {masslesschi}), so we have

\begin{equation}
\phi (\eta) = \frac {H_0}{k\sqrt {2}} ( i - k \eta ) ~e^{-ik \eta}~.
\end{equation}

\noindent
Up to a constant, this mode agrees with Eq. (53) in Ref. \cite {Lyth}.

At late times, $\chi_2$ diverges, while $\phi$ freezes. The latter property
has often been used in early studies as a heuristic argument for the 
scale invariant nature of fluctuations in an inflationary universe. The 
expectation value for the Heisenberg operator $\phi^2$ will also approach
a constant. In the Schroedinger picture this property translates into the freezing out of the width  for the one-mode wave function. Therefore, the 
\lq\lq rolling down the hill\rq\rq $~$ in the $\chi$ representation, which is 
responsible for the growth of the particle number, is equivalent in the 
massless 
case to the Lyth's criteria (in $\phi$ representation) of a wave packet 
which does not spread appreciably. The two criteria for classical behavior are 
in agreement.

\vfil

\newpage

\end{document}